\documentclass[conference]{IEEEtran}

\usepackage{cite}
\usepackage{array}
\newcolumntype{C}[1]{>{\centering\arraybackslash}p{#1}}
\usepackage{amsmath,amssymb,amsfonts}
\usepackage{algorithmic}
\usepackage{graphicx}
\usepackage{caption}
\usepackage{subfigure}
\usepackage{booktabs} 
\usepackage{multirow}
\usepackage{algorithm}
\usepackage{xcolor}

\usepackage[papersize={8.5in,11in}, left=.8in, right=.87in, top=.8in, bottom=1.01in]{geometry}
\def\BibTeX{{\rm B\kern-.05em{\sc i\kern-.025em b}\kern-.08em
    T\kern-.1667em\lower.7ex\hbox{E}\kern-.125emX}}
\begin{document}

\title{ Erasing Radio Frequency Fingerprints via Active Adversarial Perturbation\\}

\author{\IEEEauthorblockN{Zhaoyi Lu\IEEEauthorrefmark{1},
Wenchao Xu\IEEEauthorrefmark{2}, 
Ming Tu\IEEEauthorrefmark{1},
Xin Xie\IEEEauthorrefmark{1},  
Cunqing Hua\IEEEauthorrefmark{1},
Nan Cheng\IEEEauthorrefmark{3}
}

\IEEEauthorblockA{\IEEEauthorrefmark{1}School of Cyber Science and Engineering, Shanghai Jiao Tong University, Shanghai, China,\\
\IEEEauthorrefmark{2}Department of Computing, The Hong Kong Polytechnic University, Hong Kong, China,\\
\IEEEauthorblockA{\IEEEauthorrefmark{3}School of Telecom. Engineering Xidian University, Shaanxi, China}
}

\IEEEauthorblockA{luzhaoyi@sjtu.edu.cn,
wenchao.xu@polyu.edu.hk,
tonymtu@sjtu.edu.cn,\\
xiexin\_312@sjtu.edu.cn,
cqhua@sjtu.edu.cn,
nancheng@xidian.edu.cn
}
 }

\maketitle

\begin{abstract}

Radio Frequency (RF) fingerprinting is to identify a wireless device from its uniqueness of the analog circuitry or hardware imperfections. However, unlike the MAC address which can be modified, such hardware feature is inevitable for the signal emitted to air, which can possibly reveal device whereabouts, e.g., a sniffer can use a pre-trained model to identify a nearby device when receiving its signal. Such fingerprint may expose critical private information, e.g., the associated upper-layer applications or the end-user. In this paper, we propose to erase such RF feature for wireless devices, which can prevent fingerprinting by actively perturbation from the signal perspective. Specifically, we consider a common RF fingerprinting scenario, where machine learning models are trained from pilot signal data for identification. A novel adversarial attack solution is designed to generate proper perturbations, whereby the perturbed pilot signal can hide the hardware feature and misclassify the model. We theoretically show that the perturbation would not affect the communication function within a tolerable perturbation threshold. We also implement the pilot signal fingerprinting and the proposed perturbation process in a practical LTE system. Extensive experiment results demonstrate that the RF fingerprints can be effectively erased to protect the user privacy.

\end{abstract}


\section{Introduction}

Wireless radio frequency (RF) fingerprinting has emerged as an promising solution for physical layer identification, which is leveraged for supporting a variety of security applications, such as anti-impersonation, intrusion detection, etc. \cite{sankhe2019no}  The principle behind RF fingerprinting is that unlike soft ID, e.g., MAC address, the hardware imperfections of wireless devices caused during the manufacturing process cannot be spoofed using other radio interfaces, and thus can be regard as a reliable hardware identification, e.g., applied as a non-cryptographic technique for authentication \cite{soltanieh2020review}. 

Typical RF fingerprinting process include signal capturing, feature mining and classification \cite{liu2021machine}. To capture the unique features from the emitted signal, the frequency offset \cite{brik2008wireless}, transient pattern \cite{danev2010attacks}, I-Q imbalance, power amplifier non-linearity \cite{cekic2021wireless}, etc., are leveraged to determine the device signature, which can be differentiated even for devices of the same deign and product line \cite{polak2015wireless}. In recent years, machine learning models are widely applied to exploit the hidden feature from the signal data \cite{hanna2022wisig}, i.e., perform supervised training for classifying the device categories \cite{ezuma2019micro}, or using unsupervised algorithm to cluster similar fingerprints to infer the device manufacturers \cite{huang2023wyner}.

However, as a coin has two side, such inevitable feature from the emitted signal of a device can also expose its presence, i.e., a base station could identify the device by comparing its unique signal distortions with previously memorized fingerprints, or use a pre-trained model to infer the device information such as device manufacturers, RF category, etc.
With the proliferation of personal smart devices and the emerging large pre-trained models with strong generalization capability, it can be foreseen that such RF fingerprint can provide confirmatory information about the device \cite{mohammed2023radio}, which can substantially expose user's privacy, such as user location, network activity, etc. For example, by recognizing the device within the coverage area of a base station, the up-layer application or even the real end-user associated to the device can be located and tracked, which can be taken advantage by malicious purposes, e.g., detecting user behavior or direct ads, etc. 

\begin{figure*}[htbp]
    \centering
    \includegraphics[width=1.0\linewidth]{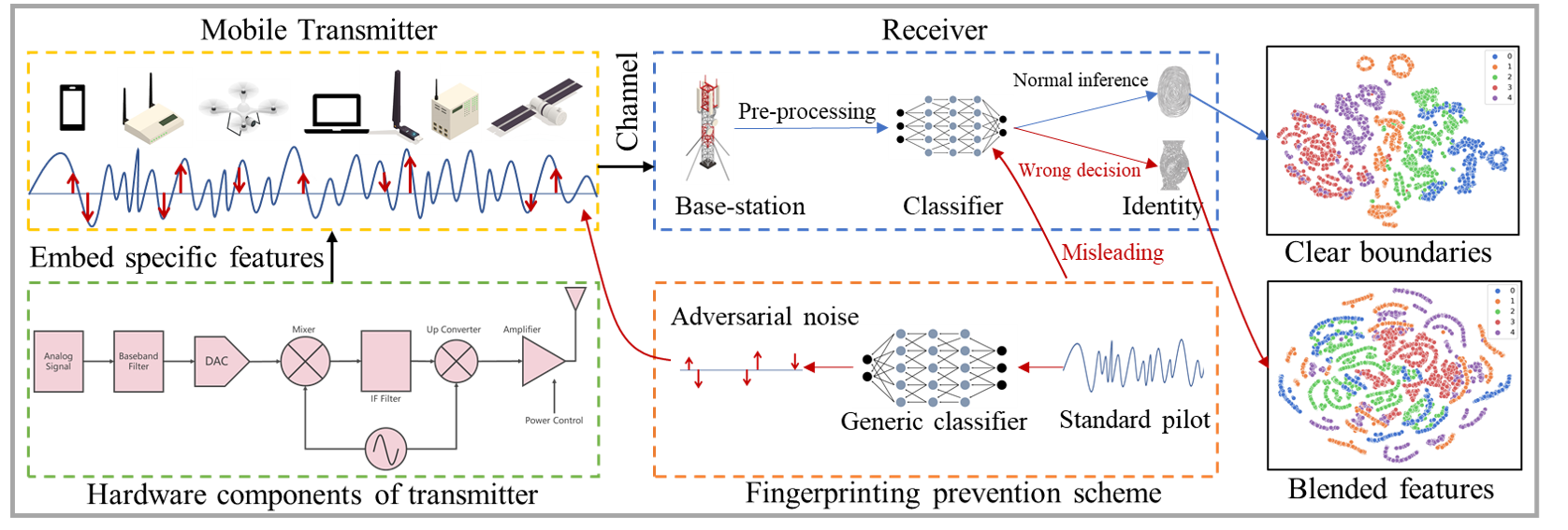}
    \caption{ Illustration of erasing RF fingerprint via active adversarial perturbation in detail. }
    \label{fig:pipeline}
\end{figure*}

To deal with the above issue, in contrast to existing literature that dedicate to utilize the RF fingerprinting for identification, this paper considers to decouple the relationship between the signal feature and the device hardware, i.e., to erase the fingerprints from the signal perspective, and hide those hardware imperfections so that they could no longer be inferred by machine learning models trained over previous signal data. Our solution is to generate artificial perturbations to the original signal, whereby the impairment caused from hardware imperfections would no longer contribute to the classification output of the model. As shown in the visualization of the feature locations of the original and the perturbed signal data in Fig. \ref{fig:pipeline}, the decision boundaries is clear and thus the model can classify the fingerprints to corresponding ID. While after perturbation, the decision boundaries cannot be discerned, i.e., the identification can be concealed. The main challenges are threefold. First, the perturbation should be able to misclassify the model to wrong identifications. Secondly, the perturbed signal should not affect the normal communication function, i.e., maintaining stable packet delivery and link rate with limited perturbation noise budget. In addition, to avoid misleading the channel estimation function, the perturbation noise's efficiency deserves scrutiny, i.e., to minimize the noise budget as much as possible when perturbing the signal.

Specifically, we consider a widely-used fingerprinting scenario, where the stationary signals is fingerprinted for identification, e.g., preamble, training symbol, and pilot signal, etc. We apply the adversarial attack mechanism to hide the specific features targeting a pre-trained model. In order to set a proper threshold of the perturbation noise power, we theoretically show that there is an upper bound to ensure the communication process would not be affected. Furthermore, a heuristic noise power control method is designed to optimise the perturbation efficiency. 

In order to validate the proposed methodology, a practical fingerprinting system was implemented within 4G LTE networks. Furthermore, we demonstrate that the model trained on the original signal is unable to identify any device after perturbation, despite the communication system functioning correctly. From the perspective of the signal, it is possible to erase the fingerprint. The paper's contributions are as follows.

\begin{figure}[htbp]
    \centering
    \includegraphics[width=1.0\linewidth]{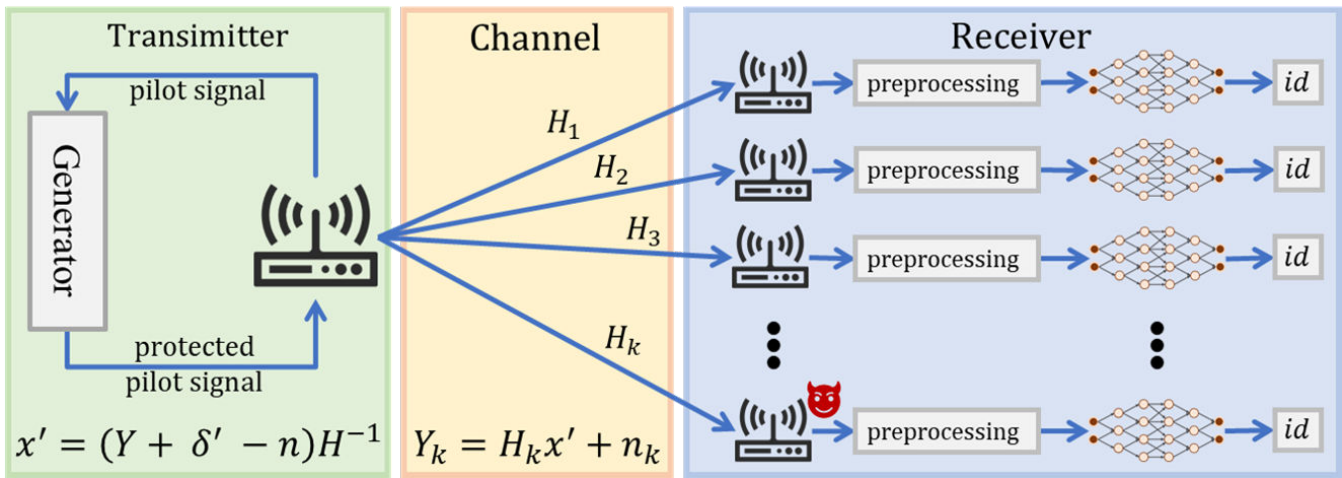}
    \caption{The pipeline of the RF fingerprinting erasure. There are three phases during the process: generation of the protected pilot signal, transmission, and receiving. We design algorithms for generator to add protective noise.  }
    \label{fig:pipeline2}
\end{figure}

\begin{itemize}

\item To the best of our knowledge, we are the first to consider erasing the fingerprint by adversarial perturbation to the signal emitted to air to decouple the specific signal feature with the device.

\item We theoretically analyze the artificial perturbations on the channel functions, which can derive the upper bound of the noise budget.

\item We conduct extensive experiments in a practical LTE network, and the results demonstrate that our method can effectively erase the fingerprint while not affecting the communication process.

\end{itemize}

\section{Preliminaries}

\noindent\textbf{LTE protocol and Pilot signal:}
LTE protocol adopts the orthogonal frequency-division multiplexing (OFDM) access technology, which is the  mainstream access technology utilized in most of the mobile radio standards\cite{8044342}.
The OFDM system utilizes the equidistant arrangement of pilot symbols in a lattice structure to implement the Minimum Mean-Squared Error (MMSE) channel estimation. Besides, the diamond arrangement in the time-frequency plane achieves the optimal estimation in the condition of a uniform pilot symbol grid\cite{8045816}. 
The pilot signal is designed to estimate the channel when demodulating, and its position has been specified in the protocol. Furthermore, the bits content of the pilot signal is known for users. Compared with the data part, the pilot signal is a fixed sequence that subtle changes among devices could be captured to identifying. And the above characteristics provide the possibility to build a pilot-based fingerprinting scheme, and it is now widely used.

\noindent\textbf{Transferability:}
Fingerprint erasing depends on the effective adversarial noise, which is generated according to the network based classifier. If the general classification model has been acquired in advance, the adversarial noise can be injected using the white-box method. However, if an unknown model is deployed for fingerprinting, it is almost impossible for the transmitter to acquire the detail. Fortunately, researchers have demonstrated that there is transferability between networks in the same task. This implies that the adversarial samples can not only affect the original model, but also can affect another one, even if the architectures and the datasets used for training are different \cite{goodfellow2014explaining, papernot2016transferability, liu2016delving}. Consequently, it is possible to generate adversarial noise with the substitute model in order to attack the target model.

\noindent\textbf{Trade-off:}
The core idea of our method is to mislead the fingerprinting model at the receiver end without severe communication degradation. Here, we leverage the adversarial noise to accommodate the fingerprinting erasure purpose. However, the artificial noise brings the negative effect on the demodulation process, because the additional noise of the pilot signal increase the channel estimation error, which is vital for the following process.

\section{System model}

The RF fingerprint could be generally obtained through three phases: signal acquisition, pre-processing, and fingerprinting identification. 
Specifically, the baseband signal $s(t)$ will firstly go through the hardware components, such as the power amplifier, which embeds the specific hardware features into the wireless signal. Let us denote the function of hardware components $f(\cdot)$, and the baseband signal is denoted by $x(t) = f(s(t))$. $Y$ represents the received signal at the receiver end, which is given by:
\begin{equation}
    Y = Hx+n,
\label{eq:baseband_signal}
\end{equation}
where $H$ is the estimated channel, and $n$ refers to the channel noise. After pre-processing of  filtering, normalization, and synchronization, the signal could be extracted. Then the pre-trained classifier model at the receiver end implements the identification classifying, making decision based on the subtle signal distortion and the prior knowledge of the model. The prediction process of the classifier is as follows:
\begin{equation}
    \hat{y} = l(Y,\theta)
\label{eq:prediction}
\end{equation}
where  $\hat{y}$ denotes the decision of the classifier, $l$ represents the model classifier, and $\theta$ is the model parameters.

In order to hide the fingerprint of the RF devices, it is supposed to perturb the signal with tiny perturbation so that the classifier model could be misled. Let $Y^{\prime}$ denote the perturbed signal, which is given by:
\begin{equation}
    Y^{\prime} = Y+\delta,
\label{eq:add_perturbation}
\end{equation}
where $\delta$ represents the designed perturbation, which can be generated with the FGSM strategy as follows ($\epsilon$ is the perturbation budget): 
\begin{equation}
    \delta = \epsilon\cdot{sign}(\nabla_Y{L(\theta,Y,y)}),
\label{eq:fgsm}
\end{equation}

As a result, the decision of the classifier model could be changed to $\hat{y}^{\prime}$ as follows: 
\begin{equation}
    \hat{y}^{\prime} = l(Y^{\prime},\theta),
\label{eq:new_decision}
\end{equation}

Further, we consider to make the artificial noise sparse enough to control the average noise power. Here, we propose the power-controlled fingerprinting prevention strategy. In equation (\ref{eq:sparsification}), $\delta^{\prime}$ denotes the sparsified perturbation, $g$ represents the sparsification function, and $r$ is the perturbing ratio, which denotes the 
proportion of the perturbed pilot signal elements.
Besides, the power of artificial noise is constrained with the hyper-parameter $s$, and the proper range of $s$ could be derived from the experiments and analysis.
\begin{equation}
    \delta^{\prime} = g(\delta, r, s)
\label{eq:sparsification}
\end{equation}

Then the defingerprinting signal $x^{\prime}$ can be obtained with the estimated channel $H$ at the receiver end:
\begin{equation}
    x^{\prime} = (Y+\delta^{'}-n)H^{-1}
\end{equation}

The paper focus on protecting the device identity privacy, thus ensuring the system usability is the premise. There are two popular metrics to evaluate the communication performance: the Packet Loss Rate and the Block Error Rate (BLER). Further, the packet loss rate may not increase even if the Block Error Rate grows, that is because the communication system has the fault tolerance mechanism. 
Therefore, the BLER could be regarded as the more advanced  restrictions to ensure the communication quality. Accordingly, it is expected to implement the fingerprinting erasure with as little impact as possible, namely trying to minimize the BLER metric when the perturbation is introduced for fingerprinting prevention ($\mathop{\min} BLER$). In other words, in order to maintain the performance of communication functions (BLER is acceptable), the noise power must be constrained within a proper range ($\text{s.t.}\ \sigma_{pert}^2 \leq s$).

\section{Algorithm Design}
This section outlines the design of effective methods to erase the RF fingerprint. The first strategy is to introduce adversarial noise into the pilot signal. Furthermore, the algorithm is enhanced with a sparse prevention method, which mitigates the adverse impact on the pilot signal by reducing the average power of the artificial noise, preserving the communication performance. The details are provided below.

\subsection{Straightway Fingerprinting Prevention}
The straightway fingerprinting prevention employs the artificial noise generated by the FGSM adversarial method, whereby the perturbation is added to the signal in the opposite direction of gradient descent. 
It is assumed that the model used to generate the adversarial noise can be acquired (white-box attack). If the model is unknown to the transmitter, a substitute model can be employed to generate effective adversarial noise due to the transferability of the model in the same task.

\begin{algorithm}[htbp]
\caption{Straightway Fingerprinting Prevention}
\label{algo:1}
\begin{algorithmic}[1]
\REQUIRE Model parameters $\theta$, standard pilot signal $Y$.
\ENSURE Protected pilot signal $Y^{\prime}$ \\

\STATE Initialization of the perturbation budget $\epsilon$.\\
\STATE Inference $\hat{y} = l(Y,\theta)$.\\
\STATE Get the loss and conduct the back-propagation.\\
\STATE Obtain the input gradients $\nabla_{Y} {J(\theta,Y,y)}$.\\
\STATE Perturb $Y^{\prime}=Y+\epsilon\cdot{sign}(\nabla_{YJ(\theta,Y,y)})$.\\
\STATE Return protected pilot signal $Y^{'}$.\\

\end{algorithmic}
\end{algorithm}

Algorithm~\ref{algo:1} shows the details of the proposed method. Firstly, the initialization is conducted to configure the noise power (budget $\epsilon$), which is selected within the proper range given by the analysis and experiments. Then the generator could obtain the gradients of the pilot signal input from the classifier model with back-propagation. Furthermore, the perturbation is added to the original pilot signal sequence. In this instance, the label utilised for inference is set as the number of the target device, which may lead the classifier to make erroneous decisions.

\subsection{Power-controlled Fingerprinting Prevention}

\begin{algorithm}[htbp]
\caption{Power-controlled Fingerprinting Prevention}
\label{algo:2}
\begin{algorithmic}[1]
\REQUIRE Model parameters $\theta$, standard pilot signal $Y$.
\ENSURE Protected pilot signal $Y^{\prime}$ \\
\STATE Initialization of the budget $\epsilon$ and the perturbing ratio $r$.
\STATE Inference $\hat{y} = l(Y,\theta)$.\\
\STATE Get the loss and conduct the back-propagation.\\
\STATE Obtain the input gradients $\nabla_{Y_{i,j}}{J(\theta,Y,y)}$.\\
\STATE Sort the input elements.\\
\STATE Acquire influential elements (r$\%$).\\
\STATE Perturb $Y_{i,j}^{\prime}=Y_{i,j}+\epsilon\cdot{sign}(\nabla_{Y_{i, j}}{J(\theta,Y,y)})$.\\
\STATE Return protected pilot signal $Y^{'}$.\\

\end{algorithmic}
\end{algorithm}

The ratio of perturbation to budget and the perturbation budget itself influence the noise power and may degrade the channel estimation accuracy. Therefore, it is necessary to maximize the efficiency of the noise, which can be achieved by using as little noise as possible. Our method is improved with sparse adversarial perturbation, which is generated based on the contributions to the classifier decision. Specifically, only the pilot signal that contributed to the classifier decision is injected with the perturbation, rather than indiscriminately perturbing all the pilot data samples.

Similar with the Algorithm~\ref{algo:1}, the initialization of the hyper-parameters, the inference of the input sample and the back-propagation are necessary (steps 1, 2 and 3). In step 4, the gradients are obtained individually as $Y_{i, j}$ to denote the contributions of the element of $i$ row and $j$ column of the input sample. Further, the pilot signal elements will be sorted according to the gradients (step 5). The perturbing ratio $r$ and the noise power $\epsilon$ could be decided within the range given by analysis and experiments. Thus, the influential elements would be perturbed for fingerprinting prevention, and the protected pilot signal can be utilized for communication. 

\begin{figure}[htbp]
\centering  
\subfigure[]{
\label{fig:SNR_loss}
\includegraphics[width=0.45\linewidth]{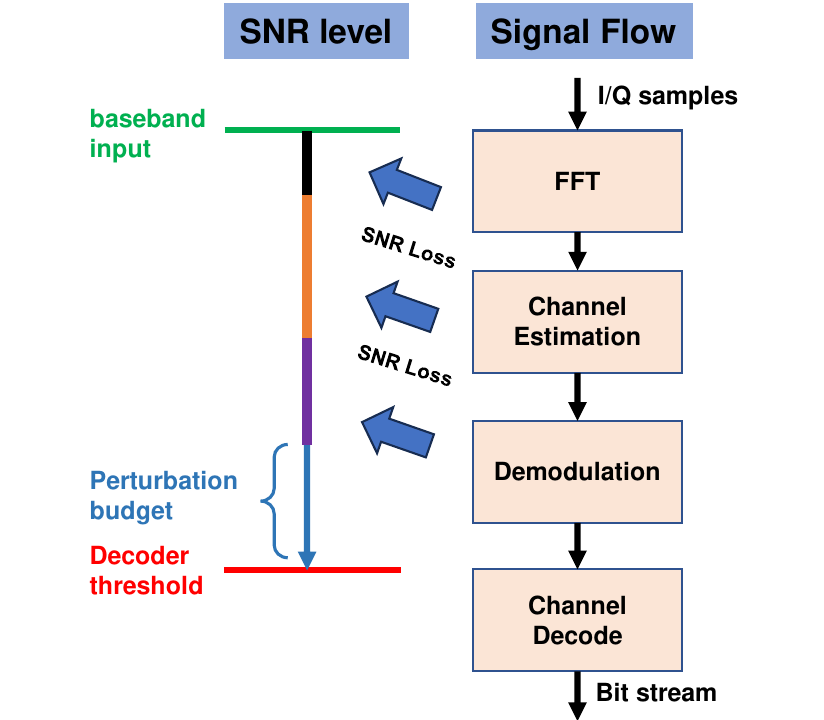}}
\subfigure[]{
\label{fig:devices}
\includegraphics[width=0.45\linewidth]{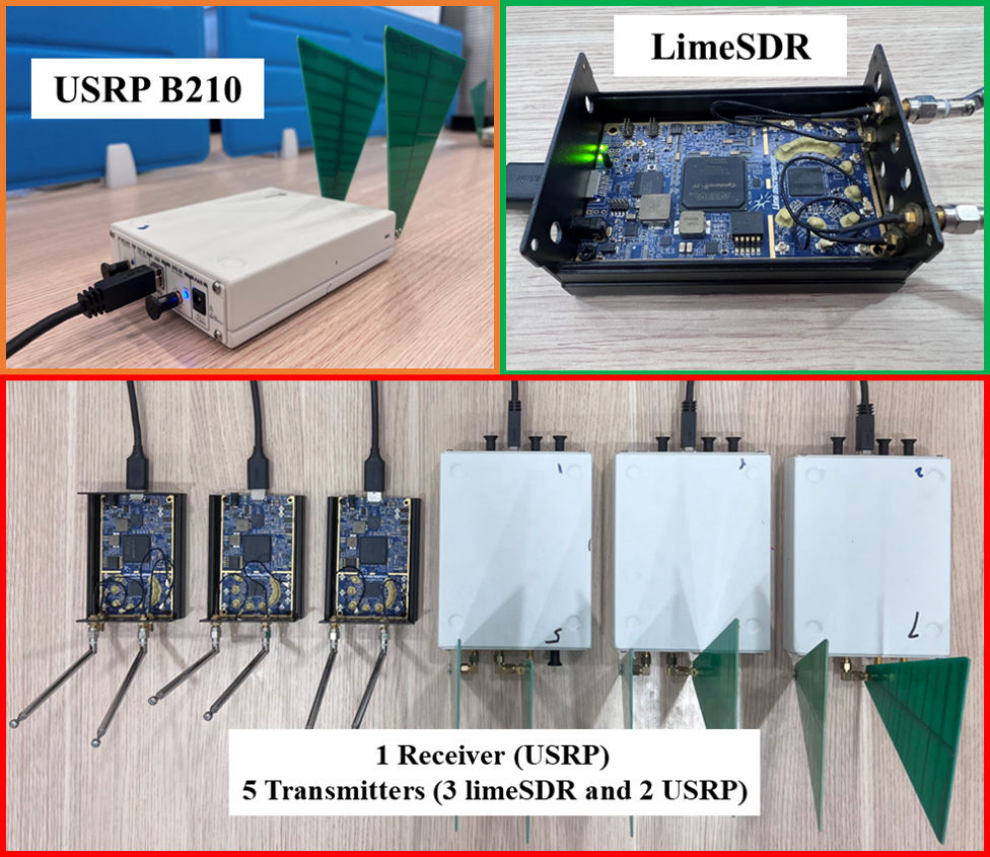}}
\caption{SNR budget for perturbation and SDR devices used in the experiments.}
\label{fig:SNR_loss&devices}
\end{figure}

\section{Implementation}
In this section, we first introduce the configuration of the SDR-based testbed, the dataset generation, and the classifier model.
Then, the metric, the hyperparameters and the evaluation strategy are presented.

\noindent\textbf {Setup.} 
The testbed consists of 3 USRP B210 and 3 LimeSDR USB. One of the USRPs acts as a receiver and the other 2 USRPs and 3 LimeSDRs act as transmitters with different identities. These devices are driven by srsRAN, an open source 4G/5G platform consisting of complete UE/eNodeB protocol stacks and a lightweight Core Network (CN) protocol stack, running on a personal computer supported by Intel Core i7-13700H cpu and Ubuntu 20.04 system. 

\noindent\textbf {Dataset.} 
Considering the hardware characteristic is changing with the environment conditions, such as the temperature, distance and standby time, we collect the dataset in 10 different conditions. In every condition, the transmitters communicate with the receiver and the pilot signals are extracted for one-batch dataset. Specifically, each batch contains 1000 data samples, which is extracted from 10 sub-frames in one frame. The dataset has been split into training set and the test set with the splitting ratio of 0.2.

\noindent\textbf{Models and Training.}
We designed a convolutional neural network (CNN) to implement the classifier. The neural network has five convolutional layers including three (3, 2) kernel size and two (3, 1) kernel size convolutional layers. In addition, each convolutional layer uses a ReLU function and a maximum pooling layer. We then configured three fully connected layers.
We trained the model using an SGD optimiser with a learning rate of ${10}^{-3}$, and the model can achieve 99\% accuracy on the test set.

\noindent\textbf{Metrics.}
To conduct a comprehensive evaluation, we adopt metrics to evaluate the impact on communication and the protection performance. 

\begin{itemize}

\item{Communication performance:}
Three classic metrics are considered, including the Block Error Rate, the Bit Rate and the Packet Loss Rate (PLR).

\item{Protection performance:}
To demonstrate the effectiveness of the designed scheme, the protection success rate (PSR) is proposed.  When the classifier model makes a wrong decision about the device identity, the current disturbance is considered as a valid sample. And the PSR is measured from the quotient of the valid sample number and the total sample.
     
\end{itemize}

\noindent\textbf{Hyperparameters.}
There are two hyperparameters that can be adjusted in the experiment, including perturbing ratio and the perturbation budget, which refer to the proportion of signals that are perturbed, and the average noise power per symbol perturbed, respectively.

\noindent\textbf{How to evaluate.}
The well-trained classifier model is considered as the target fingerprint identifier, which is deployed at the receiver side. Before transmission, we first generate the effective perturbation based on the model with proposed scheme. The artificial perturbation is then injected into the pilot, allowing the destruction of specific features. After reception, the received signal is fed into the classifier after the necessary processing, and if the decision is incoherent with the real identity of the device, the protection is considered successful.

\section{Performance  Evaluation}
\begin{figure}[htbp]
    \centering
    \includegraphics[width=0.9\linewidth]{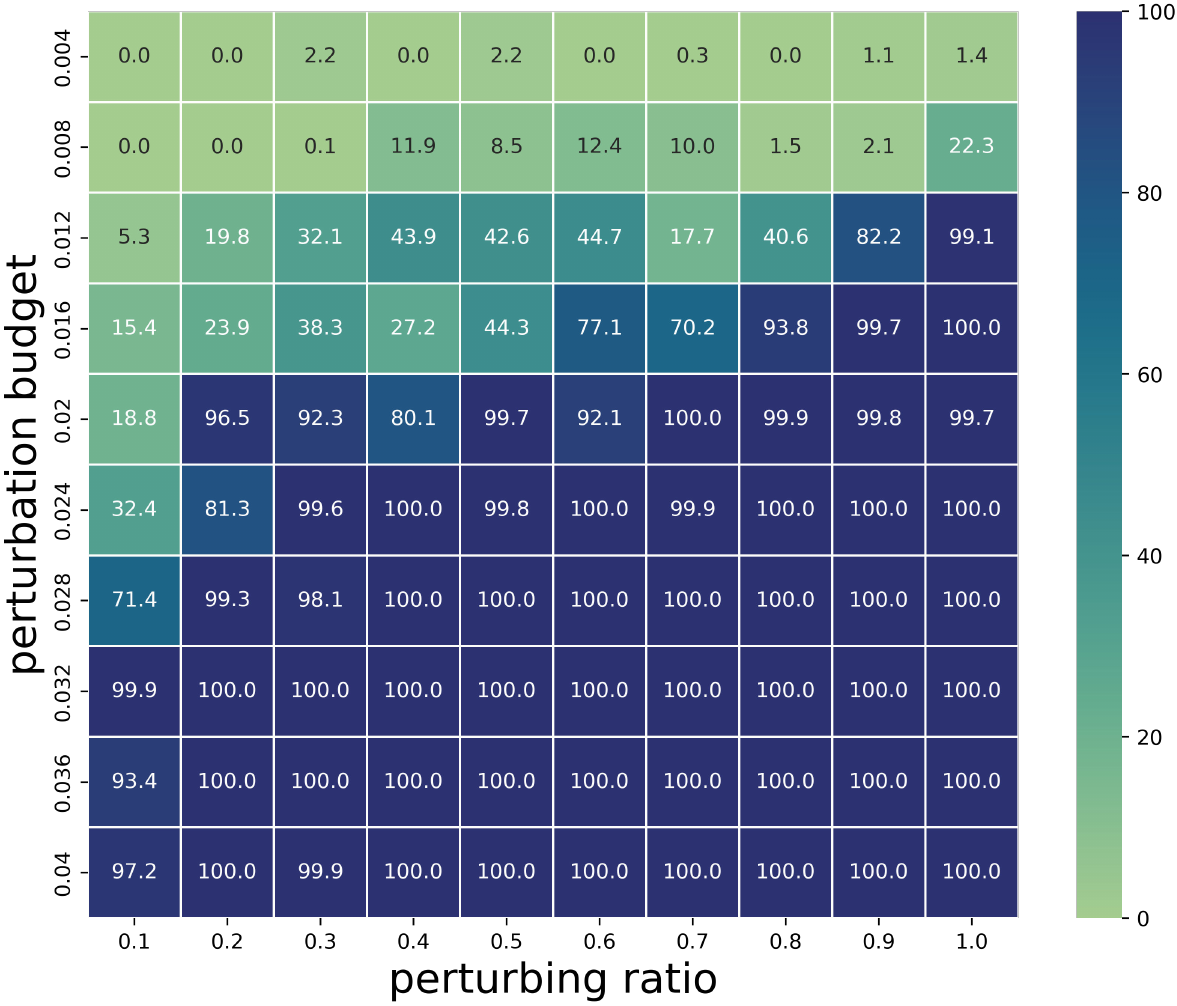}
    \caption{Heatmap of the protection success rate (\%) results with different ratios and budgets.}
    \label{fig:results1}
    \vspace{-1em}
\end{figure}

\begin{figure}[htbp]
\centering  
\subfigure[PSR among ratios]{
\label{fig:ratios}
\includegraphics[width=0.45\linewidth]{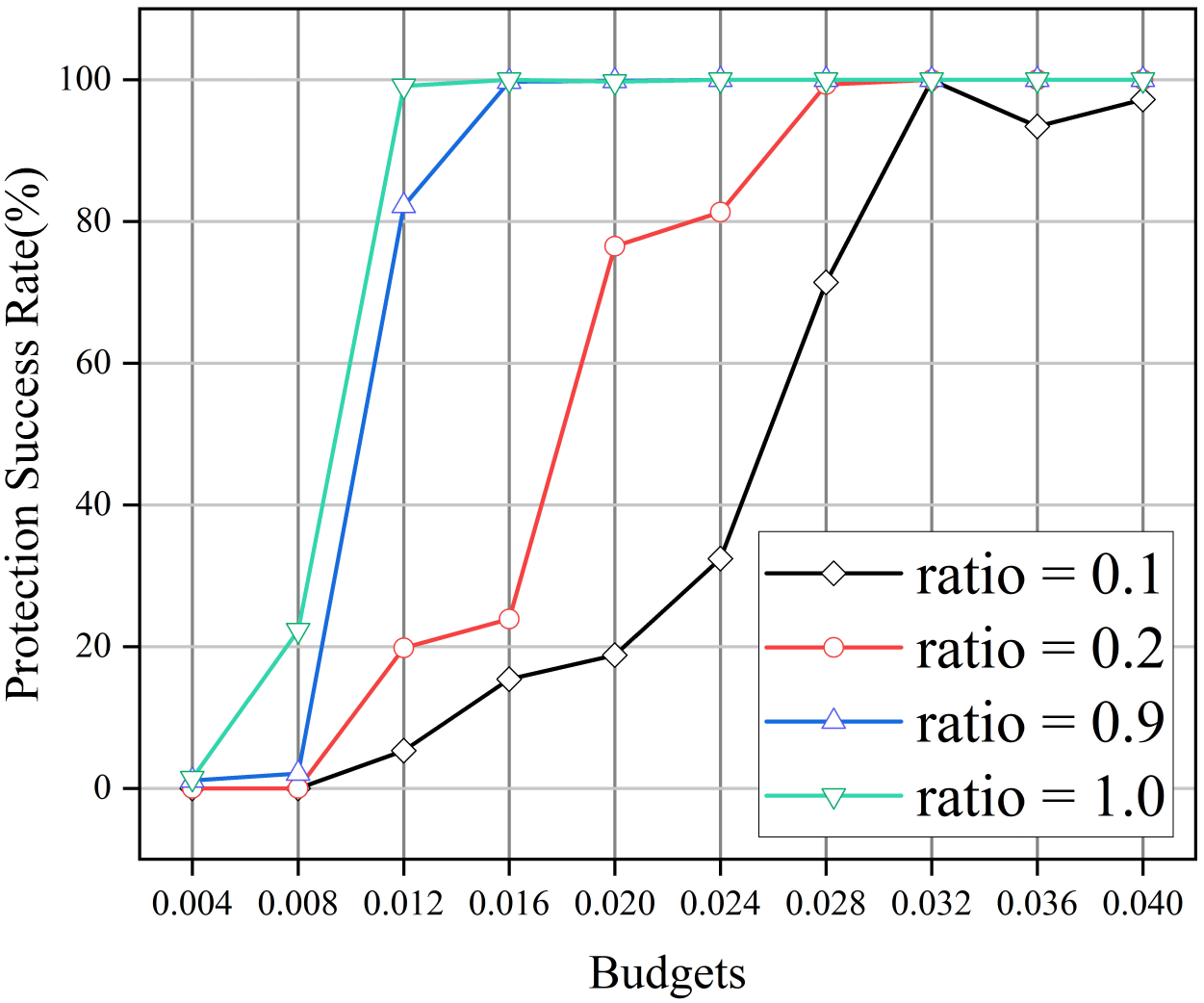}}
\subfigure[Ablation experiment]{
\label{fig:random}
\includegraphics[width=0.45\linewidth]{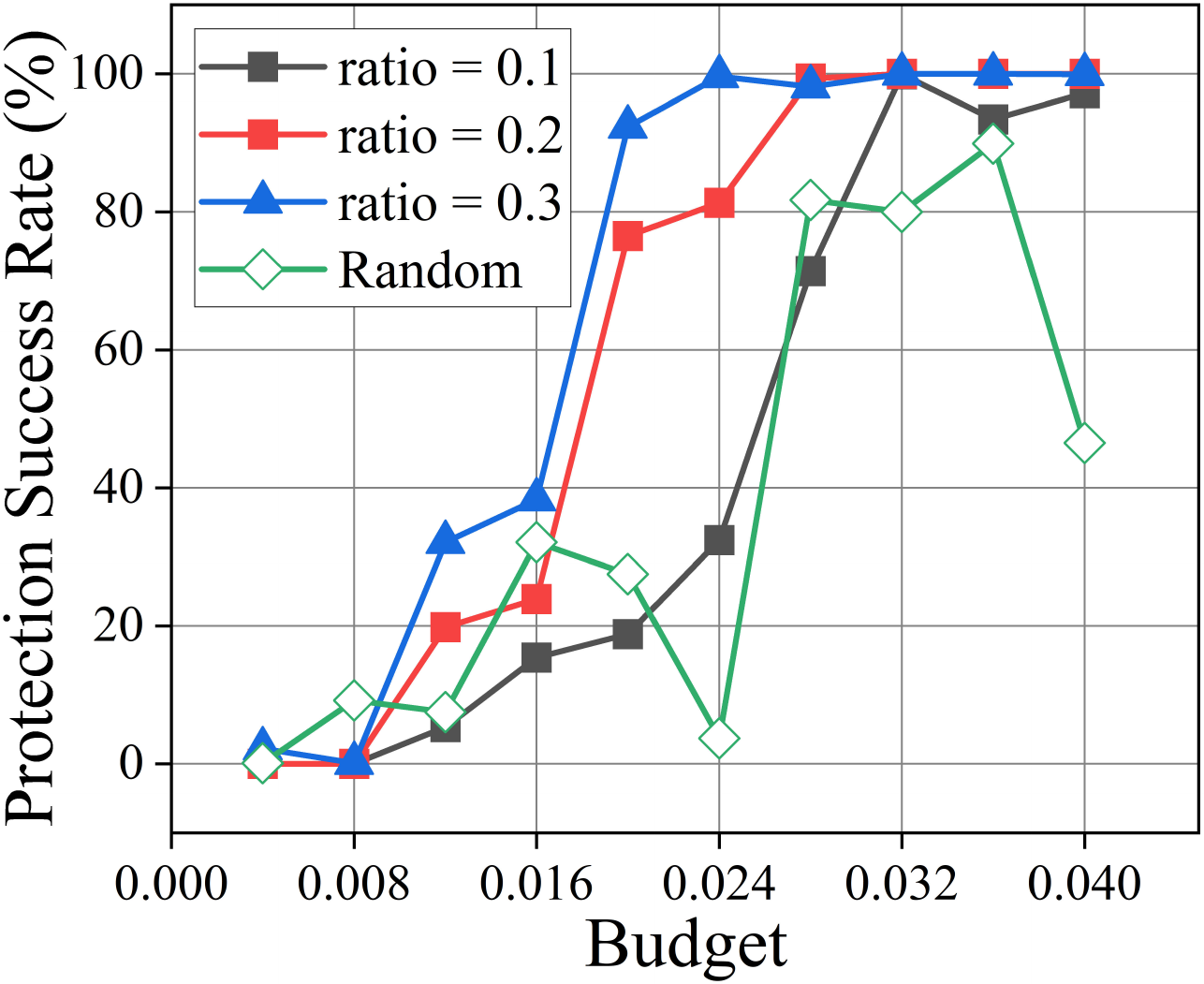}}
\caption{\ref{fig:ratios}-Comparison of the protection success rate versus ratios. \ref{fig:random}-Ablation experiment results to compare the random noise and the generated noise.}
\label{fig:ratios&ablation}
\end{figure}

\noindent\textbf{Protection success rate.}
We conduct extensive experiments with the SDR devices, and the results are illustrated as follow. The heat map in Fig.~\ref{fig:results1} shows the protection success rate of the power-controlled method (Algorithm~\ref{algo:2}). Two hyper-parameters are considered during the experiments that the perturbing ratio and the perturbation budget varies within the proper range. The perturbing ratio denotes the proportion of the perturbed pilot signal elements, and 1.0 represents to perturb all the elements, corresponding to the straightway method in Algorithm~\ref{algo:1}. The above results illustrate that the proposed method can erase the fingerprint even if the perturbing ratio is at 0.1, while most of the pilot signal elements remain as they are. Fig.~\ref{fig:ratios} compares the protection success rate among different perturbing ratio: the PSR increases faster when given a larger ratio, which could also compensate the smaller budget. Besides, the BLER remains 0 under all the conditions during experiments. And this result proves that the effective hyper-parameters are far away from the threshold of communication degradation.

\begin{figure*}[t]
\centering

\subfigure[Bitrate of TCP]{
    \label{fig:bitrate_tcp}
    \includegraphics[width=0.22\linewidth]{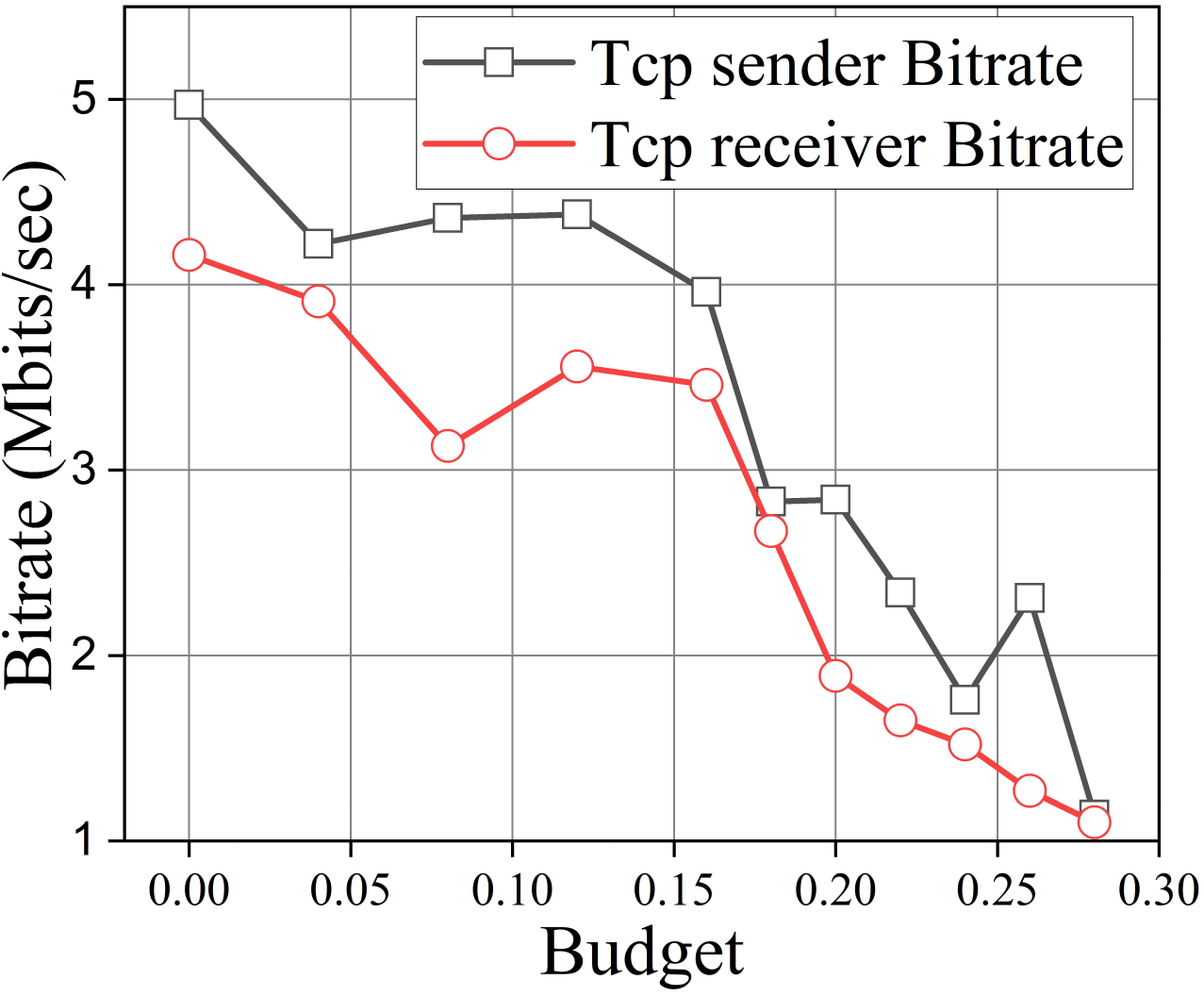}}
\subfigure[Bitrate of UDP]{
    \label{fig:bitrate_udp}
    \includegraphics[width=0.22\linewidth]{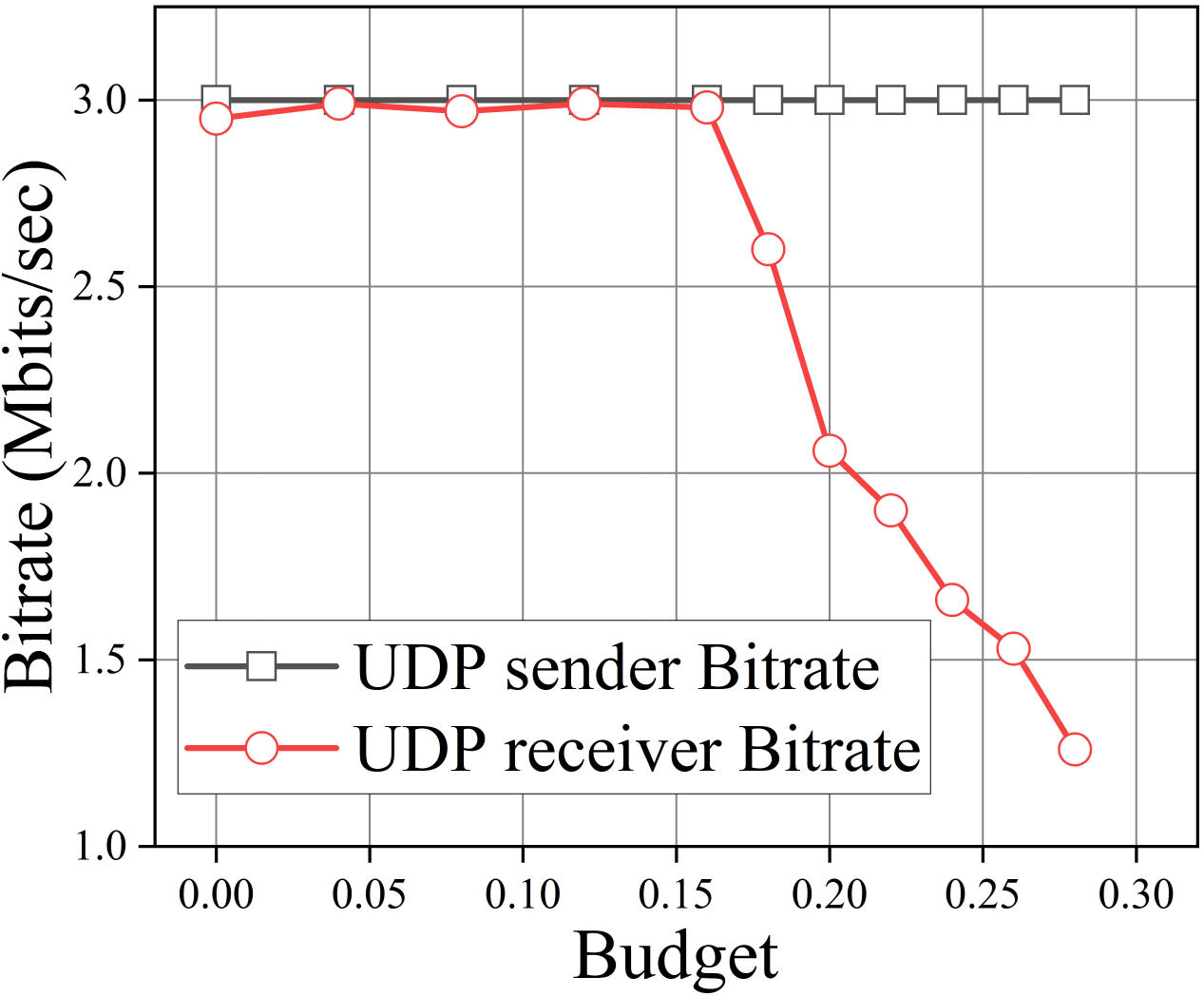}}
\subfigure[Packet loss rate of UDP]{
    \label{fig:plr_udp}
    \includegraphics[width=0.22\linewidth]{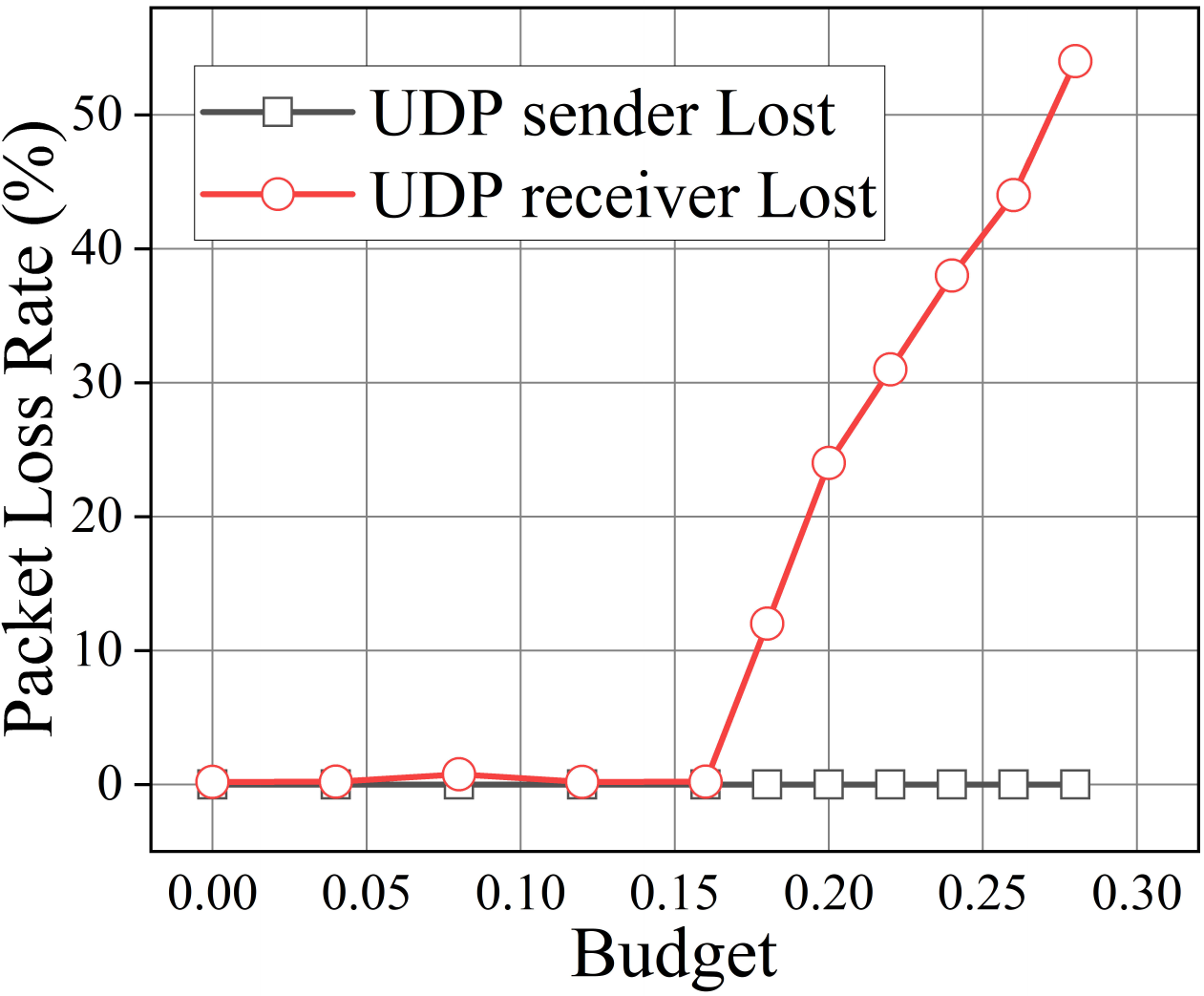}}
\subfigure[BLER]{
    \label{fig:bler}
    \includegraphics[width=0.22\linewidth]{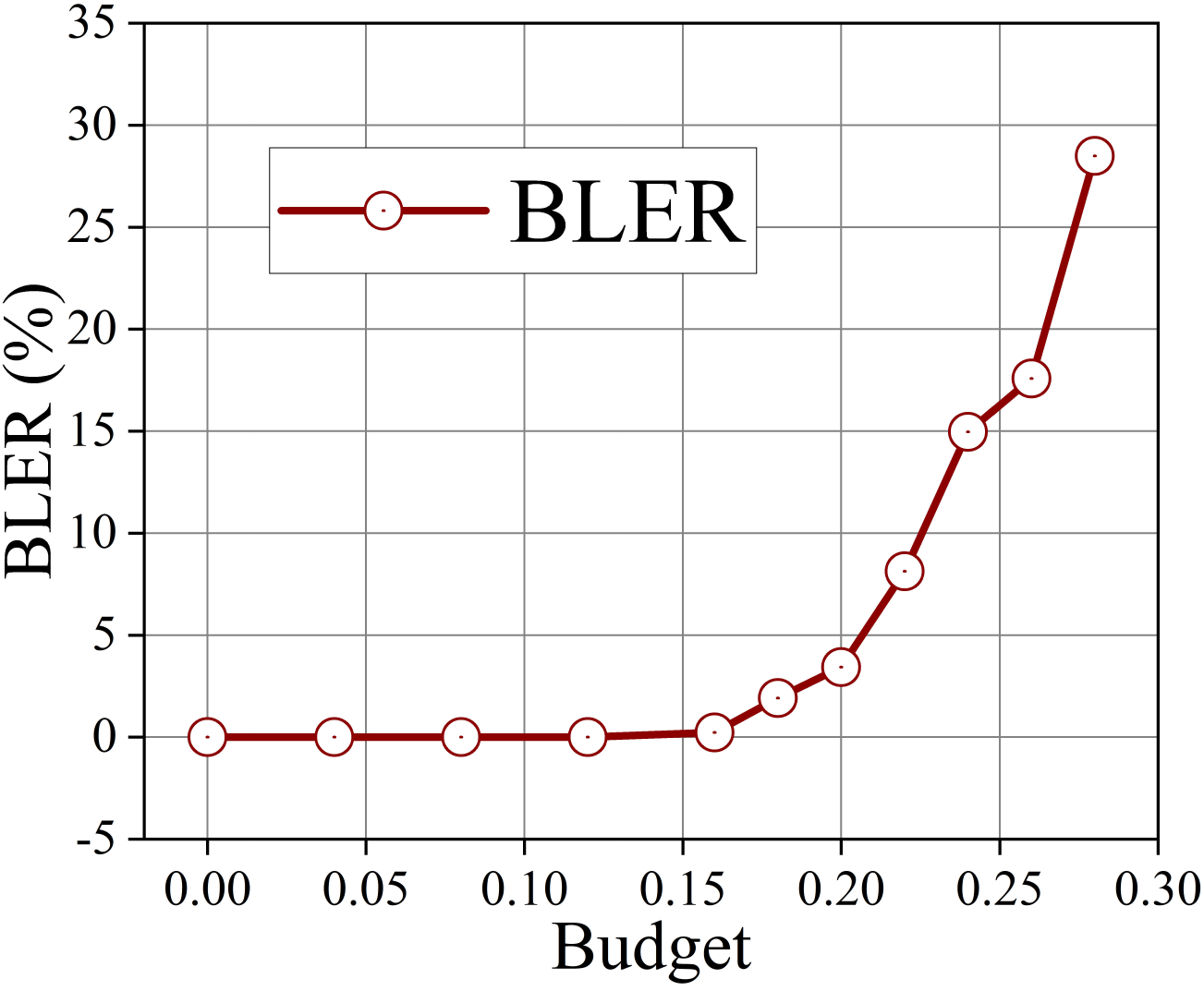}}

\caption{Fig.~\ref{fig:bitrate_tcp} and \ref{fig:bitrate_udp} show the TCP and UDP Bitrate (Mbits/sec). The bandwidth is set at 3Mbps for UDP and unlimited for TCP. Fig.~\ref{fig:plr_udp} and \ref{fig:bler} illustrate the UDP packet loss rate and the BLER versus budgets.}
\label{fig:TCP_UDP}
\end{figure*}

\noindent\textbf{Impact on communication.}
Furthermore, we compare the communication quality under different conditions to investigate the feasibility of deployment, and the UDP and TCP protocols are considered. 
Fig.~\ref{fig:bitrate_tcp} and \ref{fig:bitrate_udp} show the bitrate curves during the communication process. We expect to observe the impact on the communication quality under different power of artificial perturbation. In this experiment, the perturbing ratio is configured at 1.0, and the budget varies from 0 to 0.28. It is evident that the bitrate significantly decreases when the budget is bigger than 0.16. The bitrate at the receiver end is acceptable when the budget is smaller than the threshold of 0.12, three times bigger than the range configured in our experiment (the biggest budget in our experiment is at 0.04). Thus, the degradation caused by the proposed method could even be ignored. Furthermore, Fig.~\ref{fig:plr_udp} and \ref{fig:bler} illustrates the results of the UDP packet loss rate and the BLER with different budgets. Both curves maintain at 0 when the budget is smaller than 0.16. These results provide a consistent conclusion with the bitrate curves.

Note that Fig.~\ref{fig:TCP_UDP} and Fig.~\ref{fig:results1} are not drawn on the same scale, it is because we want to show that there is the margin to tolerate the artificial noise that the maximum noise power adopted in our experiments ($\epsilon = 0.04$) is also well below the threshold at which communication degrades ($\epsilon = 0.16$).

\noindent\textbf{Ablation experiment.}
Further, an ablation experiment was conducted to compare the erasing effectiveness between the proposed method and random noise. Random noise was generated with a random gradient sign, whereby each pilot element to be perturbed was injected with a random noise of expected power (corresponding to the perturbation budget). The perturbation ratio of the random noise was set to 1.0, as in previous experiments. Fig.~\ref{fig:random} shows that random noise is unable to perform as well as the proposed method. The fingerprint prevention performance of the random noise is inferior to that of the proposed method, with a ratio of 0.2, in terms of the same budget. Furthermore, the performance is unstable. In conclusion, the proposed method can achieve more effective and stealthy protection.


\section{Discussion}
\noindent\textbf{Extensibility:} The proposed scheme is not only compatible with the 4G LTE protocol, but also with WiFi and Bluetooth protocols, among others. Most neural network-based RF fingerprinting systems are designed to utilise stationary signals, including the preamble, training symbol, pilot signal and reference signal, among others. These types of signals are designed to perform appropriate communication functions, such as channel estimation. Consequently, the method can be extended to other protocols.

\noindent\textbf{Effect on communication:}
The artificial perturbation is injected onto the pilot signal, which is designed to conduct channel estimation. The error caused by the channel estimation would influence the demodulation performance. Although our studies have shown that there is a limited effect on communication, we will further investigate this issue in more detail.


\section{Conclusion}
This paper presents a study of the potential for erasing the fingerprints of wireless devices via active RF signal perturbation. An adversarial attack mechanism is proposed for generating the perturbations to the pilot signal, which is widely used for fingerprinting tasks. In order to enhance the stealthiness and reduce the impact on communication, a power-controlled prevention strategy is put forth, which can further optimize the perturbation injection strategy through the selection of perturbation position. The extensive experiments demonstrate that the proposed scheme is highly effective in achieving fingerprint hiding without compromising communication performance. In the future, we will further investigate the impact of perturbation on communication and extend the scheme to other protocols.


\bibliography{biblio}

\begin{thebibliography}{10}
\providecommand{\url}[1]{#1}
\csname url@samestyle\endcsname
\providecommand{\newblock}{\relax}
\providecommand{\bibinfo}[2]{#2}
\providecommand{\BIBentrySTDinterwordspacing}{\spaceskip=0pt\relax}
\providecommand{\BIBentryALTinterwordstretchfactor}{4}
\providecommand{\BIBentryALTinterwordspacing}{\spaceskip=\fontdimen2\font plus
\BIBentryALTinterwordstretchfactor\fontdimen3\font minus \fontdimen4\font\relax}
\providecommand{\BIBforeignlanguage}[2]{{%
\expandafter\ifx\csname l@#1\endcsname\relax
\typeout{** WARNING: IEEEtran.bst: No hyphenation pattern has been}%
\typeout{** loaded for the language `#1'. Using the pattern for}%
\typeout{** the default language instead.}%
\else
\language=\csname l@#1\endcsname
\fi
#2}}
\providecommand{\BIBdecl}{\relax}
\BIBdecl

\bibitem{sankhe2019no}
K.~Sankhe, M.~Belgiovine, F.~Zhou, L.~Angioloni, F.~Restuccia, S.~D’Oro, T.~Melodia, S.~Ioannidis, and K.~Chowdhury, ``No radio left behind: Radio fingerprinting through deep learning of physical-layer hardware impairments,'' \emph{IEEE Transactions on Cognitive Communications and Networking}, vol.~6, no.~1, pp. 165--178, 2019.

\bibitem{soltanieh2020review}
N.~Soltanieh, Y.~Norouzi, Y.~Yang, and N.~C. Karmakar, ``A review of radio frequency fingerprinting techniques,'' \emph{IEEE Journal of Radio Frequency Identification}, vol.~4, no.~3, pp. 222--233, 2020.

\bibitem{liu2021machine}
Y.~Liu, J.~Wang, J.~Li, S.~Niu, and H.~Song, ``Machine learning for the detection and identification of internet of things devices: A survey,'' \emph{IEEE Internet of Things Journal}, vol.~9, no.~1, pp. 298--320, 2021.

\bibitem{brik2008wireless}
V.~Brik, S.~Banerjee, M.~Gruteser, and S.~Oh, ``Wireless device identification with radiometric signatures,'' in \emph{Proceedings of the 14th ACM international conference on Mobile computing and networking}, 2008, pp. 116--127.

\bibitem{danev2010attacks}
B.~Danev, H.~Luecken, S.~Capkun, and K.~El~Defrawy, ``Attacks on physical-layer identification,'' in \emph{Proceedings of the third ACM conference on Wireless network security}, 2010, pp. 89--98.

\bibitem{cekic2021wireless}
M.~Cekic, S.~Gopalakrishnan, and U.~Madhow, ``Wireless fingerprinting via deep learning: The impact of confounding factors,'' in \emph{2021 55th Asilomar Conference on Signals, Systems, and Computers}.\hskip 1em plus 0.5em minus 0.4em\relax IEEE, 2021, pp. 677--684.

\bibitem{polak2015wireless}
A.~C. Polak and D.~L. Goeckel, ``Wireless device identification based on rf oscillator imperfections,'' \emph{IEEE Transactions on Information Forensics and Security}, vol.~10, no.~12, pp. 2492--2501, 2015.

\bibitem{hanna2022wisig}
S.~Hanna, S.~Karunaratne, and D.~Cabric, ``Wisig: A large-scale wifi signal dataset for receiver and channel agnostic rf fingerprinting,'' \emph{IEEE Access}, vol.~10, pp. 22\,808--22\,818, 2022.

\bibitem{ezuma2019micro}
M.~Ezuma, F.~Erden, C.~K. Anjinappa, O.~Ozdemir, and I.~Guvenc, ``Micro-uav detection and classification from rf fingerprints using machine learning techniques,'' in \emph{2019 IEEE Aerospace Conference}.\hskip 1em plus 0.5em minus 0.4em\relax IEEE, 2019, pp. 1--13.

\bibitem{huang2023wyner}
T.-H. Huang, T.~Dahanayaka, K.~Thilakarathna, P.~H. Leong, and H.~E. Gamal, ``The wyner variational autoencoder for unsupervised multi-layer wireless fingerprinting,'' \emph{arXiv preprint arXiv:2303.15860}, 2023.

\bibitem{mohammed2023radio}
K.~K. Mohammed, E.~I. Abd El-Latif, N.~E. El-Sayad, A.~Darwish, and A.~E. Hassanien, ``Radio frequency fingerprint-based drone identification and classification using mel spectrograms and pre-trained yamnet neural,'' \emph{Internet of Things}, p. 100879, 2023.

\bibitem{8044342}
S.~Sesia, I.~Toufik, and M.~Baker, \emph{Introduction and Background}, 2011, pp. 1--22.

\bibitem{8045816}
------, \emph{Reference Signals and Channel Estimation}, 2011, pp. 165--187.

\bibitem{goodfellow2014explaining}
I.~J. Goodfellow, J.~Shlens, and C.~Szegedy, ``Explaining and harnessing adversarial examples,'' \emph{arXiv preprint arXiv:1412.6572}, 2014.

\bibitem{papernot2016transferability}
N.~Papernot, P.~McDaniel, and I.~Goodfellow, ``Transferability in machine learning: from phenomena to black-box attacks using adversarial samples,'' \emph{arXiv preprint arXiv:1605.07277}, 2016.

\bibitem{liu2016delving}
Y.~Liu, X.~Chen, C.~Liu, and D.~Song, ``Delving into transferable adversarial examples and black-box attacks,'' \emph{arXiv preprint arXiv:1611.02770}, 2016.

\end{thebibliography}
\bibliographystyle{IEEEtran}

\end{document}